\documentclass[conference,10pt]{IEEEtran}
\IEEEoverridecommandlockouts
\usepackage{cite}
\usepackage{amsmath,amssymb,amsfonts}
\usepackage[mathscr]{eucal}
\usepackage[top=2.7cm, bottom=2.7cm, left=2cm, right=2cm] {geometry}
\usepackage{graphicx}
\usepackage{textcomp}
\usepackage{xcolor}
\usepackage{amsthm}
\usepackage{physics}
\usepackage{algorithm}
\usepackage{algpseudocode} 
\usepackage{tikz}
\usepackage[hidelinks]{hyperref}
\usepackage[switch]{lineno}
  \usetikzlibrary{arrows.meta, positioning}
\def\BibTeX{{\rm B\kern-.05em{\sc i\kern-.025em b}\kern-.08em
    T\kern-.1667em\lower.7ex\hbox{E}\kern-.125emX}}

\newtheorem{thm}{Theorem}
  \newtheorem{lem}{Lemma}

   \newcommand{\B}[1]{\mathscr{B}({#1})}
   \newcommand{\Bo}[1]{\mathscr{B}_1({#1})}

   \newcommand{\x}{\mathbf{x}}
    \newcommand{\X}{\mathbf{X}}

   \newcommand{\CH}{\mathcal{H}}
   \newcommand{\CK}{\mathcal{K}}

   \newcommand{\BC}{\mathbb{C}}

   \newcommand{\BN}{\mathbb{N}}

\let\tr\relax 
\DeclareMathOperator{\tr}{Tr}
\DeclareMathOperator{\diag}{Diag}

\begin{document}

\title{Fundamental Limits Of Quickest Change-point Detection With   Continuous-Variable Quantum States
 \\
 \thanks{This work was supported, in part, by the National Science Foundation under Grants No.~CCF-2045530 and CNS-2107265.}
 }
\author{ 
    \IEEEauthorblockN{
    Tiju Cherian John, 
        Christos N. Gagatsos, 
        and  Boulat A. Bash
    }
    \IEEEauthorblockA{
        Department of Electrical and Computer Engineering \\ 
        University of Arizona \\                               
        Tucson, Arizona, 85719, USA\\                     
        Email: tijucherian@fulbrightmail.org, \{tiju, cgagatsos, boulat\}@arizona.edu 
    }}

\maketitle
\begin{abstract}
We generalize the quantum CUSUM (QUSUM) algorithm for quickest change-point detection, analyzed in finite dimensions by Fanizza, Hirche, and Calsamiglia (Phys. Rev. Lett. 131, 020602, 2023), to infinite-dimensional quantum systems. Our analysis relies on a novel generalization of Hayashi’s theorem (Hayashi, J. Phys. A: Math. Gen. 34, 3413, 2001) concerning the asymptotics of quantum relative entropy, which we establish for the infinite-dimensional setting. This enables us to prove that the QUSUM strategy retains its asymptotic optimality, characterized by the relationship between the expected detection delay  and the average false alarm time for any pair of states with finite relative entropy. Consequently, our findings apply broadly, including continuous-variable systems (e.g., Gaussian states), facilitating the development of optimal change-point detection schemes in quantum optics and other physical platforms, and rendering experimental verification feasible. 
\end{abstract}
\section{Introduction}
The detection of abrupt changes within sequences of observations, known as change-point detection, represents a cornerstone problem in statistical analysis \cite{Wald2013-je, Siegmund2013-wx, Tartakovsky2019-pm, Tartakovsky2020-hs}. Its significance spans numerous fields, including industrial quality control, onset detection in seismic signal processing,  medical diagnostics, and environmental monitoring \cite{Basseville1993-pc}. The fundamental goal is to identify, with maximal speed and accuracy, the precise moment when the underlying statistical characteristics of observed data shift \cite{Page1954-br, Brodsky1993-wf, Lorden1971-cw}.

As quantum technologies mature, the analogous problem within the quantum domain -- detecting a change in the state or dynamics of a sequence of quantum systems -- has emerged as a vital area of research \cite{Akimoto2011-qf, Sentis2016-xj, Sentis2017-bu, Fanizza2023-ch, Mohan2023-to, Gasbarri2024-sg, Skotiniotis2024-gk, Gupta2024-cl, Guha2025-uo}, including an experimental realization in \cite{Yu2018-se}. Potential applications are foreseen in quantum sensing, monitoring quantum communication channels \cite{Guha2025-uo}, characterizing quantum processes, and diagnosing faults in quantum computing hardware \cite{Skotiniotis2024-gk}. Quantum systems, operating under the laws of quantum mechanics, present unique opportunities and challenges compared to classical systems. Notably, the possibility of employing collective measurements across multiple quantum systems offers the potential for detection strategies that outperform classical methods relying solely on individual observations \cite{Sentis2016-xj, Fanizza2023-ch}.

A critical variant of this problem focuses on \emph{quickest} change point detection \cite{Lorden1971-cw, Fanizza2023-ch, Guha2025-uo}. The objective here is not merely retrospective identification but sequential, online detection of the change with minimal delay after its occurrence, while rigorously controlling the rate of false alarms. This necessitates navigating the inherent trade-off between detection speed and reliability, a central theme in sequential analysis \cite{Tartakovsky2020-hs, Fanizza2023-ch, Gasbarri2024-sg, Guha2025-uo}.

In classical sequential analysis, the cumulative sum (CUSUM) algorithm is a widely recognized and optimal procedure for quickest change point detection \cite{Basseville1993-pc}. It operates by accumulating log-likelihood ratios and resetting the sum under certain conditions, provably minimizing detection delay for a fixed false alarm rate under standard criteria \cite{Page1954-br, Lorden1971-cw}.

Recently, Fanizza, Hirche, and Calsamiglia introduced the breakthrough idea of Quantum CUSUM (QUSUM) algorithm, which extends this paradigm to the quantum setting \cite{Fanizza2023-ch}. Their work established fundamental performance limits for quantum quickest change point detection in the asymptotic regime of long mean times between false alarms. They derived a lower bound on the achievable average detection delay and proposed measurement strategies that asymptotically meet this bound. A central finding was that the optimal trade-off is governed by the quantum relative entropy, $D(\sigma\|\rho)$, between the pre-change state $\rho$ and the post-change state $\sigma$, mirroring the role of the classical Kullback-Leibler divergence in the original CUSUM algorithm of Page \cite{Page1954-br}. In particular, as the allowed false-alarm rate vanishes (or the mean time to false alarm grows), the minimum achievable expected detection delay approaches the inverse of $D(\sigma\|\rho)$. Moreover, they proved that no quantum procedure can asymptotically outperform this bound, thereby establishing the optimality of QUSUM in the finite-dimensional regime. 

The aforementioned results in \cite{Fanizza2023-ch} rely on the assumption of finite Hilbert-space dimension. Even though systems with finite Hilbert spaces are highly relevant in quantum technology and interesting from the theory standpoint, in this work we go even further: we deliver results pertaining to continuous variables (CV) systems, i.e., systems whose quantum states are defined in infinite-dimensional spaces. Many physically relevant quantum systems require description within \emph{infinite-dimensional} Hilbert spaces \cite{Wilde2017-ti, Serafini2017-mz}. Continuous Variable (CV) quantum systems, where observables like position, momentum, or electromagnetic field quadratures have continuous spectra, are prime examples. CV systems are crucial platforms for quantum information science, underpinning protocols in quantum computation, communication, and metrology \cite{Weedbrook2012-zl, Adesso2014-sg}. Quantum optics, modeling light modes as quantum harmonic oscillators, is a canonical instance of CV systems \cite{Guha2008-iw}. Within CV systems, Gaussian states (e.g., coherent, squeezed, thermal states) are particularly important due to their experimental accessibility and mathematical tractability \cite{Ferraro2005-bh,John2021-le, Rajarama-Bhat2019-hv, John2019-io}. However, the reliance on finite-dimensional results means the QUSUM optimality analysis established in \cite{Fanizza2023-ch} does not directly apply to these systems, nor to any non-Gaussian states or other quantum systems demanding an infinite-dimensional description. This constitutes a major gap in the theory of quantum change point detection.

This paper aims to bridge this gap. 
Our main results establish the \emph{achievability} and \emph{optimality}  of the performance bound dictated by the quantum relative entropy $D(\sigma\|\rho)$, generalizing the findings of Fanizza et al. \cite{Fanizza2023-ch} to infinite dimensional states. In the process of proving achievability, we generalized a result of Hayashi   which, just like the original result in \cite{Hayashi2001-gb}, has potential applications outside of change-point detection. 
The structure of this paper is as follows. Section \ref{sec:prerequistes} introduces the change-point detection problem in our setting. In Section \ref{sec:results} we state the main results in this article and briefly mention the important ideas behind our proofs.  Section \ref{sec:CUSUM} briefly describes the CUSUM algorithm.  Section \ref{sec:achievability} presents the proof of the achievability result (Theorem \ref{thm:achievability}).  Section \ref{sec:optimality} presents the proof of the optimality result (Theorem \ref{thm:optimality}). Section \ref{sec:conclusion} concludes with a summary of our findings and discusses implications and future research directions.  Finally, in the Appendix, we  prove our generalization of the result in \cite{Hayashi2001-gb} after developing the necessary mathematical formalism.

\section{Change-Point Detection Problem }\label{sec:prerequistes}
We follow the approach introduced in \cite{Fanizza2023-ch}, adapting it to the context of infinite-dimensional states, that is, density operators in a separable Hilbert space. 
Consider a sequence of infinite-dimensional quantum states,  denoted by $\{\rho_{n}\}$ for $n\in \BN$, the set of natural numbers. This sequence models a potential change point that occurs in an unknown step $\nu+1$. Specifically, for steps $n \leq \nu$, the state is $\rho_{n} = \rho$, and for steps $n > \nu$, the state switches to $\rho_{n} = \sigma$. For notational convenience, we allow $\nu$ to take the value $\infty$, which means there is no change, that is, $\rho_n=\rho$ for all $n$.  A sequential detection algorithm interacts with sequence $\{\rho_n\}$ as follows: at each step $n$, the algorithm acquires the state $\rho_{n}$. This newly acquired state is then subjected to a joint measurement that potentially incorporates quantum information retained from states received in previous steps ($1, \dots, n-1$). The outcome of this measurement determines whether the algorithm halts at step $n$, signaling that a change is believed to have occurred, or continues to the next step $n+1$.

For a given change-point detection strategy, let $T$ denote the random variable corresponding to the step number $n$ at which the algorithm halts (the alarm time). The results of the measurement performed after receiving $\rho_{k}$ are described by discrete random variables $X_k$. We use the notation $\mathbf{X}^n = (X_1, \dots, X_n)$ for the vector of outcomes up to step $n$, and $\x^n = (x_1, \dots, x_n)$ for a specific realization of these outcomes. The decision to stop or continue at step $n$ depends on the history $\x^n$.

To analyze the performance, we define the probability measures and expectations based on the underlying state sequence. When the change point occurs at a  specific step $\nu+1\in \BN\cup\infty$, let $P_{\nu}$ represent the probability of an event, for example, the event $
[\mathbf{X}^\nu = \x^\nu$], for a specific sequence $\x^\nu$. These probabilities are generated by the sequence of quantum measurements performed by the algorithm. For the same change-point $\nu$,  let $\mathbb{E}_{\nu}$ denote the expected values given the measurement strategy, i.e., the  $\mathbb{E}_{\nu}$ is the expectation with respect to $P_{\nu}$.

A key performance metric is the mean time to a false alarm, defined as the expected stopping time when no change has occurred:
\begin{align}
\bar{T}_{\text{FA}} := \mathbb{E}_{\infty}[T].
\end{align}
Effective detection strategies should ensure that $\bar{T}_{\text{FA}}$ is large enough (for example, by exceeding a predefined threshold) to minimize premature stops ($T \leq \nu$).

In addition to false alarms, we need to quantify the algorithm's response after the change has occurred. The delay is the time elapsed between the change point and the alarm ($T-\nu$, relevant when $T > \nu$). A standard measure for this, accounting for the most challenging scenarios, is the \textit{worst-worst case mean delay}, defined in ~\cite{Fanizza2023-ch},  following~\cite{Tartakovsky2020-hs}:
\begin{align}
\bar\tau^{\star} &:=\sup_{\nu\geq 0} \sup_{\substack{\{\x^\nu |P_{\infty}[\mathbf{X}^\nu=\x^\nu]>0\}}}\mathbb{E}_{\nu}[T-\nu | T > \nu, \X^\nu=\x^\nu]. \label{eq:worst_delay} \end{align}
This metric captures the maximum expected delay, optimized over all possible change points $\nu+1$ and all possible pre-change measurement histories $\x^\nu$ (that have a non-zero probability of occurring under the no-change hypothesis), conditioned on detection occurring only after the change ($T > \nu$). It should be noted that we assume, for simplicity, the measurements with discrete outcomes in this paper.

\section{Results}\label{sec:results}
We begin this section by stating the achievability result.

\begin{thm}[\textbf{Achievability}]\label{thm:achievability}
 Let $\rho$ and $\sigma$ be two distinct density operators acting on an infinite-dimensional Hilbert space $\mathcal{H}$, representing the quantum states before and after a change point, respectively. If the quantum relative entropy $D(\sigma \| \rho) < \infty$, then for any $\epsilon > 0$, in the limit of large mean time between false alarms, $\bar{T}_{\text{FA}}$, the optimal expected detection delay $\bar\tau^\star$ achievable by a QUSUM algorithm is asymptotically bounded by:
\begin{align}\label{eq:achievability}
    \bar\tau^\star \leq \frac{\log\bar{T}_{\text{FA}}}{D(\sigma \| \rho )(1-\epsilon)} + O(1) \text{ as } \bar{T}_{\text{FA}}\to \infty.
\end{align}
\end{thm}
The formal proof of this theorem is given in Section \ref{sec:achievability}. 
The proof of Theorem \ref{thm:achievability} shows that a QUSUM strategy that employs joint measurements on blocks of incoming states can asymptotically achieve the optimal delay-false alarm trade-off characterized by $D(\sigma\|\rho)$ in \eqref{eq:achievability}. 

Except for a crucial difference our proof of Theorem \ref{thm:achievability} stated above follows similar steps as that  of the achievability result in Fanizza et al.  \cite[Theorem 1]{Fanizza2023-ch}. The difference we have is the need for a generalization of a fundamental result of Hayashi \cite{Hayashi2001-gb} on the operational attainability of the quantum relative entropy. 

Hayashi demonstrated that the quantum relative entropy $D(\rho\|\sigma)$ between finite-dimensional states $\rho$ and $\sigma$ can be asymptotically achieved as the Kullback-Leibler divergence between probability distributions obtained from a sequence of positive operator-valued measurements (POVMs) applied to $\rho^{\otimes n}$ and $\sigma^{\otimes n}$. We re-state these results for completeness in Appendix \ref{sec:Hayashi-finite-dim}. Crucially, this sequence of POVMs can be chosen depending only on one of the states ($\rho$ in our formulation below). We state our generalization of Hayashi's result in \cite{Hayashi2001-gb} to arbitrary states on infinite-dimensional separable Hilbert spaces as Lemma \ref{thm:hayashi-generalization-main-body} below. We note that, in \cite[Section 4]{Hayashi2001-gb}, Hayashi also addressed infinite-dimensional setting.  However, the result he proved is for a restricted class of states -- those states that satisfy the condition $\mu\rho\leq\sigma\leq\lambda\rho$ for some $\lambda>0$ and $\mu>0$. Many practically-important Gaussian states, including thermal states, fail to meet this requirement, which motivated our generalization. We would also like to point out that our result, just like the one in \cite{Hayashi2001-gb}, has general applications outside of quantum change-point detection.

Let us recall the measured relative entropy before stating our generalization of Hayashi's result. Given two density operators $\rho, \sigma$, and a measurement process given by a positive operator valued measurement (POVM)  $M = \{M(E)\}_{E \in \mathcal{F}}$, where $\mathcal{F}$ denotes the set of all possible outcomes of the measurements (see Appendix \ref{sec:appendix-preliminary} for more details), the measurement process yields classical probability distributions $P_{\rho,M}(E) = \text{Tr}(\rho M(E))$ and $P_{\sigma,M}(E) = \text{Tr}(\sigma M(E))$.  The measured relative entropy $D^M(\sigma\|\rho)$, with respect to $M$ is the classical Kullback-Leibler divergence between these induced distributions, $D(P_{\sigma,M} \| P_{\rho,M})$.
\begin{lem}\label{thm:hayashi-generalization-main-body}
    Let $\rho \in \Bo{\CH}$ be a density operator in an infinite-dimensional separable Hilbert space $\CH$. Then there exists a subsequence $l_n$ of natural numbers and a sequence of POVMs $\{M^n\}_{n=1}^{\infty}$ on $\CH^{\otimes{l_n}}$ such that \begin{align}
       \lim_{n\to\infty} \frac{D^{M^n}(\sigma^{\otimes{l_n}}\|\rho^{\otimes{l_n}})}{l_n} = D(\sigma||\rho), \quad \forall \sigma\in \Bo{\CH}.
    \end{align}
\end{lem}

The proof of Lemma \ref{thm:hayashi-generalization-main-body} in Appendix \ref{sec:asymptotics} consists of two steps. First, we utilize the fact that the quantum relative entropy $D(\sigma\|\rho)$ can be approximated from below by that of a sequence of finite-dimensional states $\sigma_n$ and $\rho_n$. 
 This technical lemma is stated in Appendix \ref{sec:asymptotics} as Lemma \ref{lem:algebraic} and proven subsequently. Second, we apply the finite-dimensional result of Hayashi \cite[Theorem 2]{Hayashi2001-gb} to the finite-dimensional approximation obtained in the first step.

Finally, we state the optimality result which shows that the delay-false alarm trade-off characterized by $D(\sigma\|\rho)$ that a QUSUM strategy can asymptotically achieve employing joint measurements on blocks of incoming states per Theorem \ref{thm:achievability} is, in fact, optimal.  Formally, we have:
\begin{thm}[\textbf{Optimality}]\label{thm:optimality}
Consider an algorithm designed to identify a change point between two infinite-dimensional states, $\rho$ and $\sigma$, under the condition that their relative entropy $D(\sigma||\rho)$ is finite. If this algorithm exhibits an expected false alarm time of $\bar{T}_{\text{FA}}$, its  optimal achievable delay $\bar\tau^\star$ is subject to the following lower bound for any $\epsilon > 0$: \begin{align}\label{eq:optimality}
    \bar\tau^\star\geq (1-\epsilon)\frac{\log\bar{T}_{\text{FA}}}{D(\sigma \| \rho )}(1+o(1)) \text{ as } \bar{T}_{\text{FA}}\to \infty.
\end{align}
\end{thm}

Section \ref{sec:optimality} contains the formal proof of Theorem \ref{thm:optimality}. Our proof strategy for optimality, which is typically a challenging task, is based on reducing the problem to a finite-dimensional one. This reduction is mediated by a fixed quantum channel that sends the input states to a  finite-dimensional space, enabling the use of the finite-dimensional optimality result by Fanizza et al.~\cite{Fanizza2023-ch}. By allowing all possible measurements to these finite-dimensional states, we can then invoke the result from \cite{Fanizza2023-ch} and the data processing inequality. A key aspect of our approach is the exploitation of the inverse relationship between $\bar\tau^*$ and $D(\sigma\|\rho)$, in the setting of finite dimensions which allows us to effectively combine the finite-dimensional result with the data processing inequality to prove optimality.

\section{The CUSUM (and QUSUM) Algorithms}\label{sec:CUSUM}
In this section we briefly describe the CUSUM algorithm in the classical setting. Before we begin, we note that the quantum CUSUM (QUSUM) algorithm proceeds exactly as in the classical one using the classical probability distributions obtained after performing a measurement before finally optimizing over all possible (joint) measurements on blocks of states in the history.

The Cumulative Sum (CUSUM) control chart is a powerful sequential analysis technique introduced by E.S.~Page in 1954, primarily designed for detecting changes in the underlying parameters or distribution of a process over time \cite{Page1954-br}. Unlike earlier control charts that often focused only on recent observations, CUSUM utilizes the cumulative sum of deviations or scores (frequently derived from log-likelihood ratios when distributions are known) from a target value or expected behavior. This sequential accumulation allows CUSUM to effectively leverage the entire history of the process, making it particularly sensitive to small but persistent shifts that might otherwise go unnoticed. Renowned for its computational simplicity and strong statistical optimality properties in minimizing detection delay for a given false alarm rate, CUSUM remains a fundamental and widely used tool in quality control, signal processing, finance, and various other fields requiring online monitoring \cite{Tartakovsky2019-pm, Tartakovsky2020-hs}.

The pseudocode in Algorithm \ref{alg:cusum_llr} outlines the CUSUM algorithm for sequential change detection using log-likelihood ratios. It begins by initializing the cumulative sum statistic $S$ to zero and a time counter $n$ to zero. The algorithm then enters a loop, processing observations sequentially. In each iteration, the time step $n$ is incremented, the next observation $x_n$ is acquired, and the log-likelihood ratio $\ell_n = \log(f_1(x_n)/f_0(x_n))$ is calculated, representing the evidence favoring the post-change distribution $f_1$ over the pre-change distribution $f_0$ for that observation. The core CUSUM statistic $S$ is updated using the recursive formula $S \gets \max(0, S + \ell_n)$, which accumulates the log-likelihood ratio while resetting the sum to zero if it becomes negative. Finally, the updated statistic $S$ is compared to a predefined positive threshold $h$. If $S$ reaches or exceeds $h$, the algorithm terminates and returns the current time step $n$ as the detection time, indicating that a change has been detected. Otherwise, the loop continues with the next observation.

\begin{algorithm}
\caption{CUSUM Algorithm using Log-Likelihood Ratios}
\label{alg:cusum_llr}
\begin{algorithmic}[1] 
    \Require Pre-change probability density/mass function $f_0(x)$
    \Require Post-change probability density/mass function $f_1(x)$
    \Require Detection threshold $h > 0$
    \Require Sequence of observations $x_1, x_2, \dots$

    \State Initialize CUSUM statistic $S \gets 0$ \Comment{Start sum at 0}
    \State Initialize time step $n \gets 0$

    \Loop \Comment{Process observations sequentially}
        \State $n \gets n + 1$ \Comment{Increment time step}
        \State Obtain the next observation $x_n$
        \State Calculate the log-likelihood ratio (LLR) for $x_n$:
        \State $\ell_n \gets \log \frac{f_1(x_n)}{f_0(x_n)}$ \Comment{Logarithm base is typically natural log (ln)} 
        \State Update the CUSUM statistic:
        \State $S \gets \max(0, S + \ell_n)$ \Comment{Accumulate LLR, reset to 0 if sum goes below 0} 
        \If{$S \ge h$} \Comment{Check if threshold is reached or exceeded} 
            \State \Return $n$ \Comment{Return the detection time and stop}
        \EndIf
    \EndLoop
\end{algorithmic}
\end{algorithm}
\section{Proof of Achievability (Theorem \ref{thm:achievability})}\label{sec:achievability}
 The proof of the achievability result follows exactly as in \cite{Fanizza2023-ch} except that we use Lemma \ref{thm:hayashi-generalization-main-body}  instead of the finite dimensional result of Hayashi \cite{Hayashi2001-gb} used therein. We provide a sketch of the proof here.
\begin{IEEEproof}[Proof of Theorem \ref{thm:achievability}]\label{pf:acheivability}
    Fix a POVM $M=\{M_i\}_{i\in \mathcal{I}}$ and apply it to $\rho_k$ for each $k$. An outcome $x_k$ is obtained at the $k$-the stage with probability \begin{align}\label{eq:measurement-probability}
        p(x_k)= \tr [M_{x_k}\rho] \quad\text{ or }\quad q(x_k)= \tr [M_{x_k}\sigma]
    \end{align}
    depending whether $\rho_k=\rho$ or $\rho_k=\sigma$. Note that the cardinality of the probability distributions $p$ and $q$ does not depend on the dimension of the Hilbert space $\CH$ but only on the cardinality of the POVM, $\mathcal{I}$. Hence, we can  follow the steps of the CUSUM algorithm exactly as described in \cite[Proof of Theorem 1]{Fanizza2023-ch} 
 to obtain \[\bar{\tau}^*\leq \frac{\log \bar{T}_{\text{FA}}}{D(q\|p)}+O(1), \text{ as } \bar{T}_{\text{FA}}\to \infty.\] 
    Now considering all possible measurements, define the maximal measured relative entropy \begin{align*}
        D_{\mathfrak{M}}(\sigma||\rho):= \underset{\{M|M \text{is a POVM}\}}{\sup}D^M(\sigma\|\rho)=\underset{\{\text{POVM}\}}{\sup}~D(q\|p),
    \end{align*} where the supremum is over all possible POVMs, and for each POVM,  $M=\{M_i\}$, $D_M(\sigma\|\rho)=D(q\|p)$, with $p$ and $q$ as defined in \eqref{eq:measurement-probability}. 
Generalizing beyond individual measurements, the QUSUM algorithm can utilize joint measurements on blocks of $l$ states, distinguishing between $\rho^{\otimes l}$ and $\sigma^{\otimes l}$. If the change point is assumed  a multiple of $l$, the trade-off expression modifies to:
\begin{align}\label{eq:joint-measurement}
    \bar\tau^\star \leq \frac{\log(\bar{T}_{\text{FA}}/l)}{\frac{1}{l} D_\mathfrak{M}(\sigma^{\otimes l} \| \rho^{\otimes l} )} + O(1), \text{ as } \bar{T}_{\text{FA}}\to \infty.
\end{align}
As shown in \cite{Fanizza2023-ch}, the restriction requiring the change point to be a multiple of $l$ can be lifted using methods based purely on classical CUSUM analysis. Hence, the expression \eqref{eq:joint-measurement} continues to hold in the infinte-dimensional case as well.

Next step in the proof is where our proof critically differs from that in \cite{Fanizza2023-ch}. Given $\epsilon>0$, we use our Theorem \ref{thm:hayashi-generalization-main-body} to choose an $l$ and a measurement $M^l=\{M^l_i\}_i$ depending only on $\rho$ such that \[\frac{1}{l}D(q_l\|p_l)\geq D(\sigma\|\rho)(1-\epsilon),\]
where $p_l(k) = \tr (M^l_k \rho)$ and $q_l(k) = \tr (M^l_k \sigma)$. Now \eqref{eq:joint-measurement} implies \eqref{eq:achievability} to complete the proof.
\end{IEEEproof}
\section{Proof of Optimality (Theorem \ref{thm:optimality})}\label{sec:optimality}
In the proof of our optimality result below, we define a specific subclass of measurements, $\mathbf{M}$. 
This subclass is constructed in two steps: first, we fix a quantum channel $\mathcal{M}$ that maps each incoming state $\rho_k$ to a state $\mathcal{M}(\rho_k)$ within a fixed, $n_0$-dimensional Hilbert space $\mathcal{H}_0$. 
Second, we allow any possible joint measurement to be performed on the resulting sequence of output states $\{\mathcal{M}(\rho_k)\}_{k \ge 1}$. 
Mathematically, this class is defined as
\begin{equation}\label{eq:sub-class-measurement}
    \mathbf{M} := \{M \circ \mathcal{M}^{\otimes k} \mid M \text{ is a POVM on } \mathcal{H}_0^{\otimes k}, k \in \mathbb{N} \}.
\end{equation}
Note that the dual map $\alpha:=\mathcal{M}^*$ is  a unital, normal, and CP map $\alpha: \B{\CH_0}\to \B{\CH}$. Hence, the same is true for $\left(\mathcal{M}^{\otimes k}\right)^{*}=\left(\mathcal{M}^{*}\right)^{\otimes k}$ and thus by Lemma \ref{lem:povm-under-normal-cp} $M \circ \mathcal{M}^{\otimes k}$ is a POVM on $\CH$.
\begin{IEEEproof}[Proof of Theorem \ref{thm:optimality}] Consider the sub-class $\mathbf{M}$ of measurements defined in \eqref{eq:sub-class-measurement}. Since we reduced the allowed class of measurements to a sub-class, using the fact that $A\supseteq B$ implies $\sup A\geq \sup B$, we have
 \begin{equation}\label{eq:finite-reduction}
    \bar\tau^*\geq  \bar\tau^*_{|{_\mathbf{M}}}.
\end{equation}
   Since $\mathbf{M}$ incorporates all possible measurements applied \textit{after} the fixed channel $\mathcal{M}$, the optimal expected delay achievable when restricted to this class, denoted $\bar\tau^*_{|{_\mathbf{M}}}$, is precisely equal to the optimal expected delay (or worst-worst case mean delay) for the change-point problem where the input states \textit{are} the transformed sequence $\{\mathcal{M}(\rho_k) \mid k \in \mathbb{N}\}$. So by the optimality result of Fanizza et al. \cite[Theorem 2]{Fanizza2023-ch} for any $\epsilon>0$, we have \begin{equation}\label{eq:finite-optimality}
        \bar\tau^*_{|{_\mathbf{M}}}\geq (1-\epsilon)\frac{\log\bar{T}_{\text{FA}}}{D(\mathcal{M}(\sigma) \|\mathcal{M}(\rho) )}(1+o(1)).
   \end{equation}  
   Combining \eqref{eq:finite-reduction} and \eqref{eq:finite-optimality} and using the data processing inequality we have 
   \begin{align*}
    \bar\tau^* \geq\bar\tau^*_{|{_\mathbf{M}}}&\geq(1-\epsilon)\frac{\log\bar{T}_{\text{FA}}}{D(\mathcal{M}(\sigma) \|\mathcal{M}(\rho) )}(1+o(1))\\&\geq (1-\epsilon)\frac{\log\bar{T}_{\text{FA}}}{D(\sigma \| \rho )}(1+o(1))   
   \end{align*}
   to complete the proof.
\end{IEEEproof}

\section{Conclusion}\label{sec:conclusion}
 This work extends the theoretical framework for quickest quantum change-point detection beyond its original finite-dimensional confines.
  This enables us to prove that the QUSUM strategy retains its asymptotic optimality, characterized by the relationship between the expected detection delay $\bar{\tau}^*$ and the average false alarm time $\bar{T}_{\mathrm{FA}}$:
\[
\bar{\tau}^* \sim \frac{\log \bar{T}_{\mathrm{FA}}}{D(\sigma \| \rho)} \quad \text{for large } \bar{T}_{\mathrm{FA}},
\]
for any distinct states $\rho, \sigma$ with finite relative entropy $D(\sigma \| \rho)$. 
 By deriving the ultimate performance limits applicable to general infinite-dimensional Hilbert spaces, we overcome significant mathematical and conceptual challenges inherent in continuous-variable quantum information theory. 
 
 The primary consequence of this generalization is that these fundamental bounds are now directly applicable to the broad and technologically significant class of Continuous Variable (CV) quantum states. This includes the quantum states of light that underpin much of quantum optics and form the basis for numerous quantum communication protocols. For Gaussian states $\rho$ and $\sigma$, it is known that $D(\sigma \| \rho)$ is always finite \cite{Pirandola2017-sd, Seshadreesan2018-ks,Parthasarathy2022-cv,  Androulakis2023-na,Androulakis2024-us}. Hence, the assumption $D(\sigma \| \rho)<\infty$ does not pose any challenge for them.

The results presented in this work are particularly important for optical communication and related technologies.  
Effectively, we establish the ultimate quantum limit on how quickly changes in the properties of optical signals or the channels they traverse can be detected, given a constraint on false alarms. Whether monitoring for degradation in fiber transmission \cite{Guha2025-uo}, localizing the change in transmission loss
in optical networks \cite{Zheng2025-wy}, detecting the subtle signature of an eavesdropper, identifying component malfunctions in a quantum network, or verifying the stability of quantum sensors employing CV states \cite{Wu2024-eu}, our results provide a fundamental benchmark against which practical systems can be assessed and optimized. They quantify the ultimate sensitivity afforded by quantum mechanics for this critical monitoring task in the continuous-variable domain, thereby informing the design and evaluating the potential of future quantum optical technologies. We thus solidify the theoretical foundations for sequential analysis in infinite-dimensional quantum systems and connect them directly to practical applications reliant on continuous-variable states.
\section*{Acknowledgment}
T. C. J. thanks Saikat Guha for useful discussions.  \bibliographystyle{ieeetr}
 \bibliography{change-point}
 \appendix[Asymptotics of Quantum Relative Entropy in Infinite Dimensions]
 \subsection{Hayashi's Results on the Asymptotics of Quantum Relative Entropy in Finite Dimensions}\label{sec:Hayashi-finite-dim}
 This section is dedicated to stating the main results we require from \cite{Hayashi2001-gb}. Since we want to use the order $D(\sigma\|\rho)$ in our exposition, we reformulate the results to reflect this ordering. 
\begin{thm}\label{thm:hayashi}\cite[Theorem 2]{Hayashi2001-gb}
Let $k$ be the dimension of $\mathcal{H}$ and let $\rho$ be a state on $\mathcal{H}$. Then there exists a POVM $M^n$ on the tensored space $\mathcal{H}^{\otimes n}$ which satisfies
\begin{align}\label{eq:hayashi}
D(\sigma \| \rho)-\frac{(k-1) \log (n+1)}{n} &\leq \frac{1}{n} D^{M^n}\left(\sigma^{\otimes n} \| \rho^{\otimes n}\right)\nonumber\\& \leq D(\sigma \| \rho), \quad \forall \sigma.
\end{align}
\end{thm}
The following result is not stated as such in \cite{Hayashi2001-gb} but Hayashi provides a proof of this result in Section 4 of \cite{Hayashi2001-gb}.
\begin{thm}\label{thm:Hayashi-infinite}\cite[Section 4]{Hayashi2001-gb}
Let $\rho \in \Bo{\CH}$ be a density operator in an infinite-dimensional separable Hilbert space $\CH$. Then there exists a subsequence $l_n$ of natural numbers and a sequence of POVMs $\{M^n\}_{n=1}^{\infty}$ on $\CH^{\otimes{l_n}}$ such that \begin{align}
       \lim_{n\to\infty} \frac{D^{M^n}(\sigma^{\otimes{l_n}}\|\rho^{\otimes{l_n}})}{l_n} = D(\sigma||\rho), 
    \end{align}
    for every  $\sigma\in \Bo{\CH}$ satisfying $\mu \rho\leq \sigma\leq \lambda \rho$ for some $\mu,\lambda>0$.
\end{thm}
The condition $\mu \rho\leq \sigma\leq \lambda \rho$ in the theorem above is restrictive because several important classes of states do not satisfy this condition. For example, it can be shown that two thermal states with different thermal parameters (or mean photon numbers) will not satisfy the above condition. Our novelty in Lemma \ref{thm:hayashi-generalization-main-body} is that we remove the above restriction from Hayashi's result. Nevertheless, this seemingly simple task requires some concepts from operator algebras, which we discuss in the next section.
\subsection{Operator Algebraic Preliminaries and Notations}\label{sec:appendix-preliminary}
We begin by establishing notation and recalling key definitions.  Throughout,  $\mathcal{H}$ and $\mathcal{K}$, possibly with subscripts (e.g., $\CH_i, \mathcal{K}_j$), denote complex separable Hilbert spaces, which may be finite or infinite-dimensional. The notation $\mathscr{B}(\mathcal{H})$ represents the $*$-algebra, i.e., closed under the adjoint (a.k.a. $\dagger$) operation, of all bounded linear operators on $\mathcal{H}$. The ideal generated by positive finite trace operators is called the trace-class ideal denoted by $\Bo{\CH} \left(\subseteq\mathscr{B}(\mathcal{H})\right)$. The density operators (or quantum states), which are positive semidefinite trace-class operators $\rho$ with $\text{Tr}(\rho)=1$, reside in the ideal of trace-class operators $\Bo{\CH}$. Furthermore, $\Bo{\CH}$ is a Banach space with the norm given by $\norm{T}_1=\tr (\sqrt{T^\dagger T})$. Recall that $\B{\CH}$ is isomorphic to the dual of the Banach space $\Bo{\CH}$, where $X\in \B{\CH}$ ``acts on'' $T\in \Bo{\CH}$ as \begin{align*}
    X(T) := \tr XT.
\end{align*}
In the sense of the dual relationship mentioned above, we say that $\Bo{\CH}$ is the \textit{predual} of $\B{\CH}$.

The \textit{weak topology} on $\Bo{\CH}$ generated by its dual $\B{\CH}$ is the topology in which a net $\{T_{\lambda}\}_{\lambda\in \Lambda}\subseteq \Bo{\CH}$ converges to $T\in \Bo{\CH}$ if and only if 
\begin{align*}
    \tr XT_\lambda \rightarrow\tr XT, \quad \forall X\in \B{\CH}.
\end{align*}
In this context, the \textit{weak}* \textit{topology} on $\B{\CH}$ (also known as \textit{ultra weak topology} or $\sigma$-\textit{weak topology} or \textit{normal topology})  is the topology in which a net $\{X_{\lambda}\}_{\lambda\in \Lambda} \subseteq \B{\CH}$ converges to $X\in \B{\CH}$ if and only if 
\begin{align*}
    \tr TX_\lambda \rightarrow\tr TX, \quad \forall T\in \Bo{\CH}.
\end{align*}
Two other topologies important for us in this context are weak operator topology and strong operator topology  on $\B{\CH}$. The \textit{weak operator topology} on $\B{\CH}$ is the topology in which a net $\{X_{\lambda}\}_{\lambda\in \Lambda} \subseteq \B{\CH}$ converges to $X\in \B{\CH}$ if and only if \[\mel{\xi}{X_\lambda}{\zeta}\to \mel{\xi}{X}{\zeta},\quad \forall \xi,\zeta\in \CH.\]
The \textit{strong operator topology} on $\B{\CH}$ is the topology in which a net $\{X_{\lambda}\}_{\lambda\in \Lambda} \subseteq \B{\CH}$ converges to $X\in \B{\CH}$ if and only if \[X_\lambda(\zeta)\to {X}(\zeta),\quad \forall \zeta\in \CH, \] in the Hilbert space norm.

 For a bounded with respect to the operator norm linear map $\Phi:\B{\CK}\to\B{\CH}$, there exists a unique bounded with respect to trace norm  map $\Phi_*:\Bo{\CH}\to\Bo{\CK}$,  satisfying \[\tr \left(\Phi_*(T)X \right)= \tr\left( T\Phi(X)   \right), \quad  \forall  T\in \Bo{\CH}, X\in \B{\CK}.\]  With the predual terminology described earlier we call $\Phi_*$ as the predual map of $\Phi$. A linear map $\Phi:\B{\CK}\to\B{\CH}$ is said to be \textit{completely positive} (CP) if  the map $(\text{id}_k\otimes \Phi):M_k(\BC)\otimes\B{\CK}\to M_k(\BC)\otimes\B{\CH}$ is positive (i.e., maps positive operators to positive operators), for every $k\in\BN$, where $\text{id}_k$ is the identity map on $M_k(\BC)$ satisfying  $\text{id}_k(A)=A$ for all $A\in M_k(\BC)$.   Given any bounded (with respect to the trace-norm) linear map $\Psi:\Bo{\CH}\to\Bo{\CK}$,  there exists a unique bounded (with respect to the operator-norm) linear map  $\Psi^*:\B{\CK}\to\B{\CH}$, called  the \textit{dual} of $\Psi$,  such that $\tr(\Psi(T)X)=\tr(T\Psi^*(X))$ for all $T\in\Bo{\CH}$ and $X\in\B{\CK}$. The map $\Psi^*$ is necessarily a \textit{normal} map, i.e., continuous with respect to the normal topologies on $\B{\CH}$ and $\B{\CK}$. Using the predual terminology described earlier, we see  that $\Psi$ is the predual of the map $\Psi^*$ and  $\Psi= \left(\Psi^*\right)_*$.  The map $\Psi$ is said to be CP if $\Psi^*$ is a CP map. Furthermore, $\Psi$ is \textit{trace-preserving}  (TP), i.e., $\tr(\Psi(T))=\tr(T)$ for all $T\in\Bo{\CH}$ if and only if $\Psi^*$ is \textit{unital}, i.e., $\Psi^*(I_{\CK})=I_{\CH}$, where $I_{\CK}$ and $I_{\CH}$ are the   identity maps of the respective spaces.  A trace-preserving completely positive map  $\Psi:\Bo{\CH}\to\Bo{\CK}$ is called a \textit{quantum channel}. Note that for a quantum channel $\Psi$, its dual map $\Psi^*$ is a normal map.
 
 A $*$-subalgebra of $\B{\CH}$ that contains the identity $I_\CH$ and closed under the strong operator topology is called a \textit{von Neumann algebra}.  For von Neumann algebras $\mathcal{A}$ and $\mathcal{B}$, the notation $\mathcal{A} \cong \mathcal{B}$ indicates that $\mathcal{A}$ and $\mathcal{B}$ are isometrically isomorphic as $*$-algebras, while $\iota: \mathcal{A} \hookrightarrow \mathcal{B}$ denotes an isometric $*$-embedding of $\mathcal{A}$ into $\mathcal{B}$. A linear mapping $\varphi:\mathcal{A}\to \BC$ is said to be positive if $\varphi(a^*a)\geq0$ for all $a\in \mathcal{A}$. Additionally, if $\norm{\phi} = 1$, the map $\varphi$ is called a \textit{state}. Clearly, $\B{\CH}$ is a von Neumann algebra and corresponding to a density operator $\rho\in \B{\CH}$, the map $\varphi_{\rho}$ denotes the \textit{normal state} (meaning continuous in normal topology)  on $\mathscr{B}(\mathcal{H})$ given by $\varphi_{\rho}(X) = \text{Tr}(\rho X)$ for any $X \in \mathscr{B}(\mathcal{H})$.

Araki's relative entropy, denoted $S(\omega, \varphi)$ for two normal states $\omega, \varphi$ on a von Neumann algebra $\mathcal{A}$ (typically $\mathcal{A} = \mathscr{B}(\mathcal{H})$), quantifies the distinguishability between states at the algebraic level \cite{Araki1976-ta, Araki1977-zw, Ohya1993-ku, Hiai2018-mh, Hiai2019-xn, Hiai2021-oh}. It is defined via the relative modular operator and satisfies $S(\omega, \varphi) \ge 0$, with equality holding if and only if $\omega = \varphi$.  For infinite-dimensional states represented by density operators $\rho, \rho$ in $\mathcal{H}$, the quantum relative entropy (or Umegaki relative entropy) is defined as \begin{equation}\label{def:relative-entropy-density}
    D(\sigma \| \rho):= S(\varphi_\sigma, \varphi_\rho);
\end{equation}  see  \cite[Definition 3.1, Examples 3.3 and Appendix B]{Androulakis2023-lv} and \cite{Androulakis2023-na} for the details of the definitions and an analysis of the relative modular operator in the case of states obtained from density operators. In certain special cases, for example for Gaussian sates, it is known that  $D(\sigma \| \rho) = \text{Tr}(\sigma (\log \sigma - \log \rho))$ provided that the support of $\sigma$ is contained within the support of $\rho$ ($\text{supp}(\sigma) \subseteq \text{supp}(\rho)$), and $D(\sigma \| \rho) = +\infty$ otherwise \cite{Luczak2022-ax}.

 A Positive Operator-Valued Measure (POVM), defined on a measurable space $(\Omega, \mathcal{F})$, is a map $M: \mathcal{F} \to \mathscr{B}(\mathcal{H})$ satisfying (i) positivity, i.e., $M(E) \ge 0$ for all $E \in \mathcal{F}$), (ii) countable additivity, i.e., for any countable collection $\{E_i\}_{i=1}^\infty$ of pairwise disjoint sets in $\mathcal{F}$, $M(\cup_{i=1}^\infty E_i) = \sum_{i=1}^\infty M(E_i)$, with convergence in the weak operator topology (and hence in strong operator topology as well),  and (iii) normalization, i.e., $M(\Omega) = I$, the identity operator. The outcome space $\Omega$ can be finite, countably infinite, or uncountably infinite (e.g., $\Omega = \mathbb{R}$).

Finally, the measured relative entropy connects the idea of relative entropy to measurements. Given two density operators $\rho, \sigma$ and a POVM $M = \{M(E)\}_{E \in \mathcal{F}}$, the measurement process yields classical probability distributions $P_{\rho,M}(E) = \text{Tr}(\rho M(E))$ and $P_{\sigma,M}(E) = \text{Tr}(\sigma M(E))$. Note that we used a simplified notation $p=P_{\rho,M}$ and $q=P_{\sigma,M}$ in Section \ref{sec:achievability} because the measurements in the article assume only discrete outcomes. The measured relative entropy with respect to $M$ is the classical Kullback-Leibler divergence between these induced distributions, $D(P_{\sigma,M} \| P_{\rho,M})$. Optimizing over all possible POVMs, we obtain the maximal measured relative entropy $D_{\mathfrak{M}}(\sigma \| \rho) = \sup_{M} D(P_{\sigma,M} \| P_{\rho,M})$.

\subsection{Generalization of Hayashi's Result} \label{sec:asymptotics}
We need a two technical lemmas to prove the main result of this section. We could not find Lemma \ref{lem:algebraic} in the literature, but this result might also be of independent interest.
\begin{lem}\label{lem:algebraic}
    Let $\sigma$ and $\rho$ be two states in an infinite-dimensional separable Hilbert space $\CH$. Then there is a sequence of finite-dimensional Hilbert spaces $\{\CH_n\subseteq\CH\}_{n=1}^{\infty}$ and unital CP maps $\alpha_n: \B{\CH_n}\rightarrow\B{\CH}$ such that
    \begin{align}
        D(\sigma\|\rho)= \lim_{n \to \infty}  D\left(\alpha_n{_*}(\sigma)\|\alpha_n{_*}(\rho)\right),
    \end{align}
    where $\alpha_n{_*}:\Bo{\CH}\to \Bo{\CH_n}$ is the predual map satisfying $\tr\left[X\alpha_n{_*}(T)\right]=\tr\left[\alpha_n(X)T\right] $ for all $X\in \B{\CH}$ and  $T\in \Bo{\CH}.$ In this case, we also have that the sequence on the right side of the equation above monotonically increases to the limit.
\end{lem}
\begin{IEEEproof} Figure \ref{fig:commutative-diagram} illustrates the key concept underlying the proof.    \begin{figure}
    \centering
    \begin{tikzpicture}[>=Latex, thick, auto, node distance=3cm]

  \node (A) at (0,0) {$\mathcal{A}$};
  \node (BH0) at (5,0) {$\mathscr{B}(\mathcal{H}_0)$};
  \node (BH) at (2.5,4) {$\mathscr{B}(\mathcal{H})$};
\node at (2.5,1.7) {\scalebox{1.8}{${\circlearrowright}$}};
  \draw[->] (A) -- node[left]{$\tilde{\alpha}$} (BH);
  \draw[->] (BH0) -- node[right]{$\alpha$} (BH);
  \draw[->, bend left=15] (A) to node[above]{$\iota$} (BH0);
  \draw[->, bend left=15] (BH0) to node[below]{$\delta$} (A); 

\end{tikzpicture}
    \caption{A finite dimensional von Neumann algebra $\mathcal{A}$ is isomorphic to a direct sum of full matrix algebras which is further embedded into $\B{\CH_0}$ via the inclusion map $\iota$, where $\CH_0$ is a finite dimensional subspace of $\CH$. The diagonal map $\delta$ from $\B{\CH_0}$ to $\mathcal{A}$ and the inclusion map $\iota$ are CPTP maps. Furthermore,  the map $\alpha:=\tilde{\alpha}\circ \delta$ is a CP extension of $\tilde{\alpha}$ and  also $\tilde{\alpha}=\alpha\circ\iota$. The crux of Lemma \ref{lem:algebraic} is that, by construction, $\iota$ is a recovery map for $\delta$ i.e.,  for any two states $\omega_1\circ \tilde{\alpha}$ and $\omega_2\circ \tilde{\alpha}$, where $\omega_j$ is a state on $\B{\CH}$, $j=1,2$, we have $\omega_j\circ \tilde{\alpha}\circ \delta\circ \iota = \omega_j\circ {\alpha}\circ \iota=\omega_j\circ \tilde{\alpha}$, which implies $S(\omega_1\circ\tilde{\alpha}, \omega_2\circ\tilde{\alpha}) = S(\omega_1\circ{\alpha}, \omega_2\circ{\alpha})$.}
    \label{fig:commutative-diagram}
\end{figure}
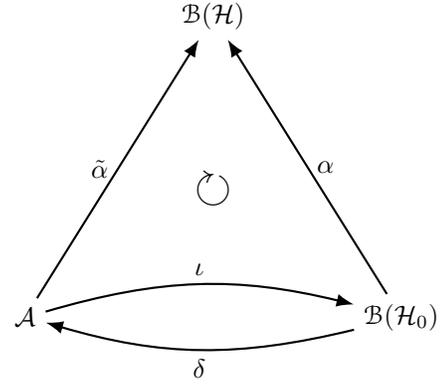
     Recall that $\B{\CH}$ is an injective von Neumann algebra \cite[Theorem 1.2.3]{Arveson1969-mv} (cf. \cite[Proposition 5.11 and Theorem 5.30]{Bryder2013-cw}). Hence by \cite[Theorem 5.30]{Ohya1993-ku} 
    \[S(\varphi_\sigma,\varphi_\rho)=\underset{\beta}{\sup}~S(\varphi_\sigma\circ\beta, \varphi_\rho\circ\beta),\]
    where $\alpha$ runs over all unital CP maps from a finite-dimensional von Neumann Algebra into $\B{\CH}$. Therefore, we get a sequence $\mathcal{A}_n$ of finite-dimensional von Neumann algebras and CP maps $\tilde{\alpha}_n:\mathcal{A}_n\rightarrow \B{\CH}$ such that \begin{align}\label{eq:ohya-petz}
        S(\varphi_\sigma,\varphi_\rho)=
    \lim_{n\to \infty}S(\varphi_\sigma\circ\tilde{\alpha}_n, \varphi_\rho\circ\tilde{\alpha}_n),
    \end{align} where the sequence on the right side of equation above is monotonically increasing to the limit.
   Recall by the structure theorem for finite-dimensional von Neumann algebras that each $\mathcal{A}_n$ is a direct sum of full matrix algebras \cite[Theorem 11.9, Chapter I]{Takesaki1979-gi}. Thus  there are integers $m(n)$ such that 
   \[\mathcal{A}_n\cong \oplus_{i=1}^{m(n)}M_{k_i}(\BC)\]
   Note also that,  we have the inclusion map $\iota_n$  sending a block diagonal matrix in $\oplus_{i=1}^{m(n)}M_{k_i}(\BC)$ to the same matrix in $M_{k_1+\cdots+k_{m(n)}}(\BC)$ identified as $\B{\oplus_{i=1}^{m(n)} \BC^{k_i}}$, \begin{align*}
      &~ \iota_n: \oplus_{i=1}^{m(n)}M_{k_i}(\BC)\to \B{\oplus_{i=1}^{m(n)} \BC^{k_i}}\\
       &a_1\oplus\cdots \oplus a_{m(n)}\underset{\iota_n}{\mapsto}\diag[a_1,\dots, a_{m(n)}]
   \end{align*}  and the identity of $\mathcal{A}_n$ is mapped to $\oplus_i I_{k_i}$ which is same as the identity of $\B{\oplus_{i=1}^{m(n)} \BC^{k_i}}$. 
  Let us denote by $\delta_n$ the diagonal map that maps a matrix $X\in \B{\oplus_{i=1}^{m(n)} \BC^{k_i}}$ to the diagonal matrix with the same diagonal entries as that of  $X$, in symbols, \begin{align*}
    &~~\delta_n: \B{\oplus_{i=1}^{m(n)} \BC^{k_i}}\rightarrow \oplus_{i=1}^{m(n)}M_{k_i}(\BC)\\
      &[x_{ij}]\underset{\delta_n}{\mapsto} \diag[x_{11},\dots,x_{m(n)m(n)}], \quad \forall n.
  \end{align*}Identifying $\oplus_{i=1}^{m(n)} \BC^{k_i}\cong{\CH}_n$, where $\CH_n$ is a $m(n)$-dimensional subspace of $\CH$, we have the following until now 
  \begin{align*}
     \scalebox{.95}{$ \mathcal{A}_n\cong \oplus_{i=1}^{m(n)}M_{k_i}(\BC)\substack{\underset{\iota_n}{\hookrightarrow}\\[-2.5em] \underset{\leftarrow} {\delta_n}}  \B{\oplus_{i=1}^{m(n)} \BC^{k_i}}\cong\B{\CH_n}{\hookrightarrow }\B{\CH}, \forall n.$}
  \end{align*}
  Note that $\delta_n$ is a unital CP map and defining $\alpha_n:= \tilde{\alpha}_n\circ \delta_n$, we get a unital CP extension of $\tilde{\alpha}_n$ to $ \B{\CH_n}$, \begin{align*}
       {\alpha}_n: & \B{\CH_n}\to \B{\CH}\\
       &X\mapsto \tilde{\alpha}_n\left(\delta_n(X)\right)\\
       &{{\alpha}_n}{_{|_{\mathcal{A}_n}}}=\tilde{\alpha}_n,\\
        \alpha_n= \tilde{\alpha}_n&\circ \delta_n \text{ and } \tilde{\alpha}_n= \alpha_n\circ\iota_n,  \quad \forall n.
   \end{align*}
  Since  $\iota_n$ and $\delta_n$ are CPTP maps,   by the data processing inequality \cite[Theorem 5.3, cf. Corollary 5.12 (iii)]{Ohya1993-ku}, we have 
\begin{IEEEeqnarray*}{rCl}
S(\varphi_\sigma\circ\tilde{\alpha}_n, \varphi_\rho\circ\tilde{\alpha}_n)&=& S(\varphi_\sigma\circ{\alpha}_n\circ \iota_n, \varphi_\rho\circ{\alpha}_n\circ \iota_n)\nonumber\\ &\leq& S(\varphi_\sigma\circ{\alpha}_n, \varphi_\rho\circ{\alpha}_n) \text{ and}\\
S(\varphi_\sigma\circ{\alpha}_n, \varphi_\rho\circ{\alpha}_n)&=& S(\varphi_\sigma\circ\tilde{\alpha}_n\circ \delta_n, \varphi_\rho\circ\tilde{\alpha}_n\circ \delta_n)\nonumber\\
&\leq & S(\varphi_\sigma\circ\tilde{\alpha}_n, \varphi_\rho\circ\tilde{\alpha}_n).
\end{IEEEeqnarray*}
   Thus we have \begin{align}\label{eq:alpha-instead-tilde-alpha}
S(\varphi_\sigma\circ\tilde{\alpha}_n,\varphi_\rho\circ\tilde{\alpha}_n )=S(\varphi_\sigma\circ{\alpha}_n, \varphi_\rho\circ{\alpha}_n).
   \end{align}
Finally, from  \eqref{eq:ohya-petz} we get
   \begin{align}\label{eq:ohya-petz-1}
        S(\varphi_\sigma,\varphi_\rho)=
    \lim_{n\to \infty}S(\varphi_\sigma\circ{\alpha}_n, \varphi_\rho\circ{\alpha}_n).
    \end{align} 
    Note that
    \begin{align}\label{eq:alpha-predual-proof}
        \varphi_{\sigma}\circ\alpha_n(X)&= \tr \sigma\alpha_n(X)= \tr \alpha_n{_*}(\sigma)X \nonumber\\
        &=\varphi_{\alpha_n{_*}(\sigma)}(X), \forall X\in \B{\CH_n}, \sigma\in \Bo{\CH}.
    \end{align}
    So we have \begin{align}\label{eq:alpha-predual}
        \varphi_{\sigma}\circ\alpha_n = \varphi_{\alpha_n{_*}(\sigma)}.
    \end{align} By definition in \eqref{def:relative-entropy-density}, and equations \eqref{eq:ohya-petz-1} and \eqref{eq:alpha-predual} we have \begin{align*}
        D(\sigma\|\rho) &=  S(\varphi_\sigma,\varphi_\rho)\\ 
        &= \lim_{n\to \infty}S(\varphi_\sigma\circ{\alpha}_n, \varphi_\rho\circ{\alpha}_n)\\
        &=\lim_{n\to \infty}S(\varphi_{\alpha_n{_*}(\sigma)}, \varphi_{\alpha_n{_*}
        (\rho)})\\
        &= \lim_{n\to \infty}D(\alpha_n{_*}
        (\sigma)\|\alpha_n{_*}
        (\rho)),
    \end{align*} where the sequence on the right side of the last line above  monotonically increases to the limit because of \eqref{eq:alpha-instead-tilde-alpha} and the fact that $\tilde{\alpha}_n$ were initially chosen in that way.
\end{IEEEproof}
\begin{lem}\label{lem:povm-under-normal-cp}
    Let $K_1$ and $K_2$ be separable Hilbert spaces and let $\alpha:\B{\CK_1}\to\B{\CK_2}$ be a unital, normal, and CP map, i.e., $\alpha_*: \Bo{\CK_2}\to \Bo{K_1}$ is a quantum channel. Let $M=\{M_i\}_{i\in \mathcal{I}}$ be a POVM on $\CK_1$. Write $\alpha(M):=\{\alpha(M_i)\}_{i\in I}$. Then $\alpha(M)$ is a POVM on $\B{\CK_2}$. Furthermore, in this case \begin{align}\label{eq:measured-rel-entropy-under-normal-cp}
D^{\alpha(M)}\left(\sigma\|\rho\right)= D^M\left(\alpha_*(\sigma)\|\alpha_*(\rho)\right) \quad \forall \sigma, \rho\in \Bo{\CK_2}.
 \end{align}
\end{lem}
\begin{IEEEproof}
    Since $\alpha$ is a positive map, $\alpha(M_i)$ is a positive operator for all $i\in \mathcal{I}$. Also, since $\alpha$ is a normal unital map we have
    \begin{align*}
\sum_i\alpha(M_i)=\alpha\left(\sum_iM_i\right)=\alpha(I)=I.
    \end{align*} Hence $\alpha(M)$ is a POVM. To prove \eqref{eq:measured-rel-entropy-under-normal-cp}, let $P=\{p_i\}_i$ and $Q=\{q_i\}_i$ where
    \begin{align*}
        p_i&= \tr \alpha(M_i)\sigma \\
        q_i&= \tr \alpha(M_i)\rho, \quad \forall i.
    \end{align*}
Since $p_i= \tr M_i\alpha_*(\sigma)$ and $q_i= \tr M_i\alpha_*(\rho)$, we see that \eqref{eq:measured-rel-entropy-under-normal-cp} is satisfied.
\end{IEEEproof}
Now we  prove Lemma \ref{thm:hayashi-generalization-main-body}. 
\begin{IEEEproof}[Proof of Lemma \ref{thm:hayashi-generalization-main-body}] We have from Lemma \ref{lem:algebraic} that there exists a sequence of finite-dimensional Hilbert spaces $\{\CH_n\subseteq\CH\}_{n=1}^{\infty}$ and unital CP maps $\alpha_n: \B{\CH_n}\rightarrow\B{\CH}$ such that
    \begin{align}\label{eq:from-lem}
        D(\sigma\|\rho)= \lim_{n \to \infty}  D\left(\alpha_n{_*}(\sigma)\|\alpha_n{_*}(\rho)\right).
    \end{align}
    Let $m(n)$ denote the dimension of $\CH_n$.
    Since $\frac{\log (x+1)}{x}\to 0$ as $x\to \infty$, we can choose an increasing sequence $l_1<l_2<\cdots $ of integers  such that
   given $n\in \BN$, \begin{align*}
       \frac{(m(n)-1) \log (l_n+1)}{l_n}< \frac{1}{n}.
   \end{align*} 
     Now by \cite[Theorem 2]{Hayashi2001-gb} (cf. Theorem \ref{thm:hayashi}) there exist a POVM $M^{l_n}$ on $\CH_n^{\otimes{l_n}}$ such that \begin{align}
0\leq D(\alpha_n{_*}\left(\sigma) \| \alpha_n{_*}(\rho)\right)&-  \frac{D^{M^{l_n}}\left(\alpha_n{_*}(\sigma)^{\otimes {l_n}} \| \alpha_n{_*}(\rho)^{\otimes {l_n}}\right)}{l_n}\nonumber\\& \leq \frac{(m(n)-1) \log (l_n+1)}{l_n} \nonumber\\
&<\frac{1}{n},\quad \forall \sigma.\label{eq:main-thm-estimate}
   \end{align}
 Let $M^n:= \alpha_n^{\otimes l_n}(M^{l_n})$ as in Lemma \ref{lem:povm-under-normal-cp}. Since $\alpha_n^{\otimes l_n}$ is a unital CP map defined on a finite-dimensional space, it is a normal map. So $M^n$ is a POVM on $\CH$. 
Given $\epsilon>0$,  choose $n$ using \eqref{eq:from-lem}  and \eqref{eq:main-thm-estimate} and  such that \begin{align*}
    \abs{\frac{D^{M^{l_n }}\left(\alpha_n{_*}(\sigma)^{\otimes l_n} \| \alpha_n{_*}(\rho)^{\otimes l_n}\right)}{l_n}-D\left(\alpha_n^*(\sigma) \| \alpha_n^*(\rho)\right)}&<\frac{\epsilon}{2}\\ \text{ and }
 \abs{D\left(\alpha_n^*(\sigma) \| \alpha_n^*(\rho)\right)-D(\sigma \| \rho)}&< \frac{\epsilon}{2}.
\end{align*}
Now by  \eqref{eq:measured-rel-entropy-under-normal-cp} and the inequalities above
\begin{align*}
&\abs{\frac{D^{M^n}\left(\sigma^{\otimes l_n} \| \rho^{\otimes l_n}\right)}{l_n}-D(\sigma \| \rho)} \\ & =\abs{\frac{D^{M^{l_n}}\left(\left(\alpha_n^{\otimes l_n}\right)_*\left(\sigma^{\otimes l_n}\right) \|\left(\alpha_n^{\otimes l_n}\right)_*\left(\rho^{\otimes l_n}\right)\right)}{l_n}- D(\sigma \| \rho)} \\
& =\abs{\frac{D^{M^{l_n }}\left(\alpha_n{_*}(\sigma)^{\otimes l_n} \| \alpha_n{_*}(\rho)^{\otimes l_n}\right)}{l_n}-D(\sigma \| \rho)} \\
&\leq \abs{\frac{D^{M^{l_n }}\left(\alpha_n{_*}(\sigma)^{\otimes l_n} \| \alpha_n{_*}(\sigma)^{\otimes l_n}\right)}{l_n}-D\left(\alpha_n^*(\rho) \| \alpha_n^*(\sigma)\right)}\\&~~~~+\abs{D\left(\alpha_n^*(\rho) \| \alpha_n^*(\sigma)\right)-D(\rho \| \sigma)} \\
&<\epsilon.
\end{align*}
This completes the proof.
\end{IEEEproof}

\end{document}